\documentclass[a4paper]{article}
\usepackage{graphicx}
\usepackage{xspace}
\usepackage[affil-it]{authblk}
\usepackage{fancyhdr}

\newcommand{\Ups}{\ensuremath{\Upsilon}\xspace}

\newcommand{\Raa}{\ensuremath{R_\mathrm{AA}}\xspace}
\newcommand{\Npart}{\ensuremath{N_\mathrm{part}}\xspace}
\newcommand{\Ncoll}{\ensuremath{N_\mathrm{coll}}\xspace}
\newcommand{\sqsn}{\ensuremath{\sqrt{s_\mathrm{NN}}}\xspace}
\newcommand{\Jpsi}{\ensuremath{J/\psi}\xspace}
\newcommand{\pT}{\ensuremath{p_\mathrm{T}}\xspace}

\title{Latest results on $\Upsilon$ production in heavy-ion collisions from the STAR experiment}

\author{R\'obert V\'ertesi (for the STAR Collaboration)\\ \texttt{vertesi.robert@wigner.mta.hu}}

\affil{Nuclear Physics Institute of the Academy of Sciences of the Czech Republic, \v{R}e\v{z} 130, 25068 Husinec-\v{R}e\v{z}, Czech Republic}
\affil{Wigner Research Centre for Physics of the Hungarian Academy of Sciences RMI, H-1525 Budapest 114, P.O.Box 49, Hungary}
\date{}

\begin{document}
\maketitle

\begin{abstract}  
We report on the latest measurements of the production of \Ups mesons in heavy-ion collisions from the STAR experiment at RHIC. New measurements of the nuclear modification factors of the \Ups{}(1S+2S+3S) and \Ups{}(1S) states in U+U collisions at $\sqsn=193$ GeV are presented as a function of the number of participants (\Npart) in the collisions. In addition, the suppression of \Ups{}(1S) and \Ups{}(2S+2S) is presented versus the quark-antiquark binding energy. Preliminary results on \Ups suppression in Au+Au collisions at $\sqsn = 200$ GeV, reconstructed via the dimuon channel, are also reported.
\end{abstract}

\section{Introduction}
There is extensive experimental evidence, eg.\ in Refs.~\cite{Adams:2005dq,Adare:2008ab}, that a hot and strongly interacting medium is formed in high-energy heavy-ion collisions. 
It is expected that color screening of the quark-antiquark binding potential within such a medium would render quarkonium formation less likely~\cite{Matsui:1986dk}, an effect colloquially referred to as "melting" of the bound state. Since quarkonium states with stronger binding energies would melt at higher temperatures than the more weakly bound states, measuring the yields of different quarkonium states may reveal a sequential suppression pattern, thus serving as a thermometer for the medium~\cite{Mocsy:2007jz}. Although the suppression of charmonia was anticipated as a key signature of the quark-gluon
plasma formation~\cite{Matsui:1986dk}, later studies revealed that several concurrent effects, such as recombination of uncorrelated $c\bar{c}$ pairs in the medium, cold nuclear matter (CNM) effects and feed-down from higher-mass states, complicate the interpretation of the results~\cite{Grandchamp:2004tn,Vogt:2012fba,Vertesi:2015xba}. Bottomonia, on the other hand, are less affected by recombination at top RHIC energies, and CNM effects are also expected to be moderate at mid-rapidity. Thus the measurements of the \Ups{}($n$S) states provide a cleaner probe of the strongly interacting medium. 

Recent results from STAR~\cite{Adamczyk:2013poh} show that in central Au+Au collisions at $\sqsn=200$ GeV, the \Ups{}(1S+2S+3S) production is suppressed to an extent that cannot be explained by CNM effects only. In the same study the yields of the excited states are consistent with a complete suppression.
Since the energy density in central U+U collisions is estimated to be about 20\% higher on the average than that in central Au+Au collisions at $\sqsn=200$ GeV~\cite{Kikola:2011zz}, studies of quarkonium production in U+U collisions can provide further tests of the sequential suppression hypothesis. Recent installation of the Muon Telescope Detector (MTD)~\cite{Ruan:2009ug}, on the other hand, opened the door for precision measurements of quarkonia at STAR via the dimuon decay channel.

\section{First measurement of \Ups{} production in U+U collisions}

\begin{figure}[h!]
\includegraphics[width=\columnwidth]{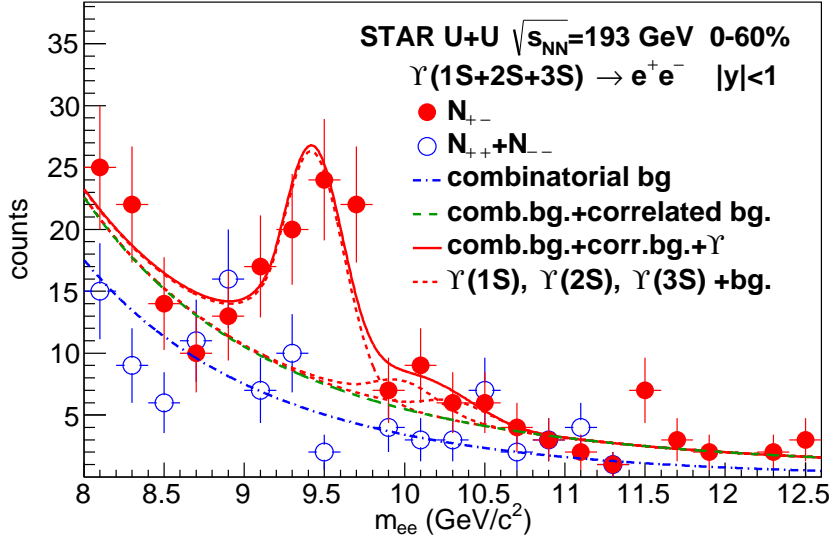}
\caption{\label{fig:uumass}Invariant mass distribution of \Ups candidates and the combinatorial background in the 0-60\% centrality class of U+U collisions at $\sqsn=193$ GeV~\cite{Adamczyk:2016dzv}.}%
\end{figure}

The STAR detector complex~\cite{Ackermann:2002ad} provides clean identification of electrons at mid-rapidity ($|\eta|<1$) in the full azimuth angle. STAR collected data in U+U collisions at $\sqsn=193$ GeV triggered by high-energy electrons (or positrons) with an integrated luminosity of 263.4 $\mu b^{-1}$. The analysis of this data, briefly recapitulated here, is described in details in Ref.~\cite{Adamczyk:2016dzv}. The tracks of the electron candidates are reconstructed in the Time Projection Chamber (TPC), and then matched with energy deposits in the Barrel Electromagnetic Calorimeter (BEMC). Electron identification is carried out in the TPC based on fractional energy loss, and in the BEMC based on the shape of the electromagnetic shower. Electron candidates have to have their energy-to-momentum ratio close to unity. Identified electron and positron candidates are then paired in order to reconstruct the invariant mass of the \Ups{} candidates. The combinatorial background is then reconstructed using like-sign electron (positron) pairs. Since Drell-Yan processes and open $b\bar{b}$ production yield correlated backround in the signal region, combinatorial background cannot be simply subtracted to get the signal. Instead, a simultaneous fit is applied on the unlike-sign and like-sign invariant mass data. The former is constructed as a sum of the three \Ups($n$S) peaks, the correlated and the combinatorial background. A priori knowledge on the shapes of the signal and background shapes were obtained from simulations~\cite{Adamczyk:2013poh,Adamczyk:2016dzv} in order to reduce the number of free parameters in the fit. 
Figure~\ref{fig:uumass} shows the invariant mass distributions together with the fitted functions on the combinatorial and correlated backgrounds as well as the signal peaks.

We use the nuclear modification factor to quantify suppression, defined as  
$\Raa=\frac{\sigma^{inel}_{pp}}{\sigma^{inel}_{AA}}\frac{1}{\langle\Ncoll\rangle}\frac{d\sigma^{AA}_\Ups
  / dy}{d\sigma^{pp}_\Ups
  / dy}$, where $\sigma^{inel}_{AA(pp)}$ is the total
inelastic cross-section of the U+U ($p$+$p$) collisions, and 
$d\sigma^{AA(pp)}_\Ups /dy$ denotes the \Ups
production cross-section in U+U ($p$+$p$) collisions.

\begin{figure}[h!]
\includegraphics[width=\columnwidth]{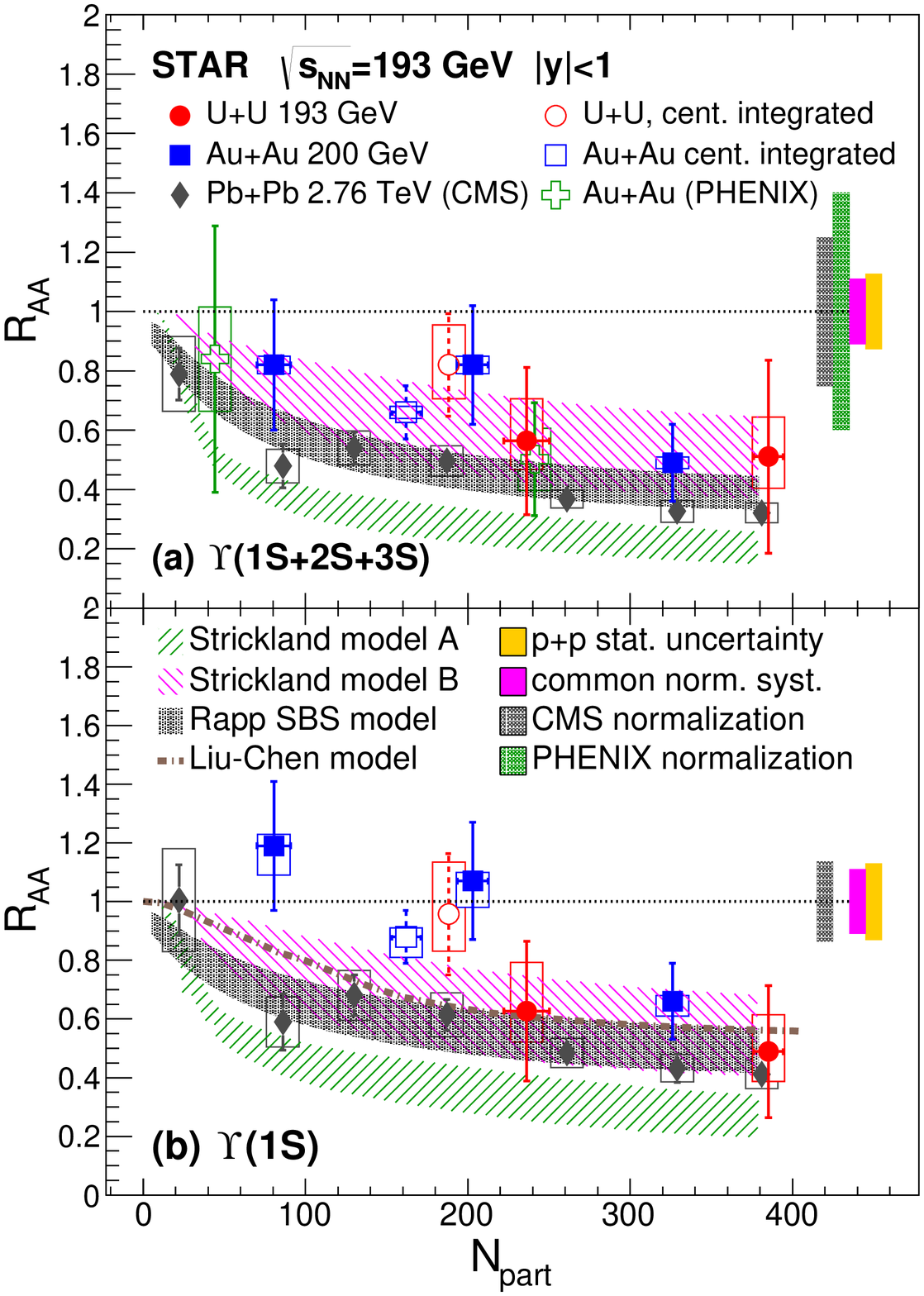}
\caption{\label{fig:raa_npart}\Raa of (a) \Ups{}(1S+2S+3S) and (b) \Ups{}(1S) as a function of \Npart in U+U and Au+Au collisions from STAR (solid circles and squares, respectively)~\cite{Adamczyk:2013poh,Adamczyk:2016dzv}, compared to PHENIX~\cite{Adare:2014hje} (crosses) and CMS~\cite{Khachatryan:2010zg} data (diamonds) as well as several theoretical calculations~\cite{Emerick:2011xu,Strickland:2011aa,Liu:2010ej}.}
\end{figure}

\begin{figure}[h!]
\includegraphics[width=\columnwidth]{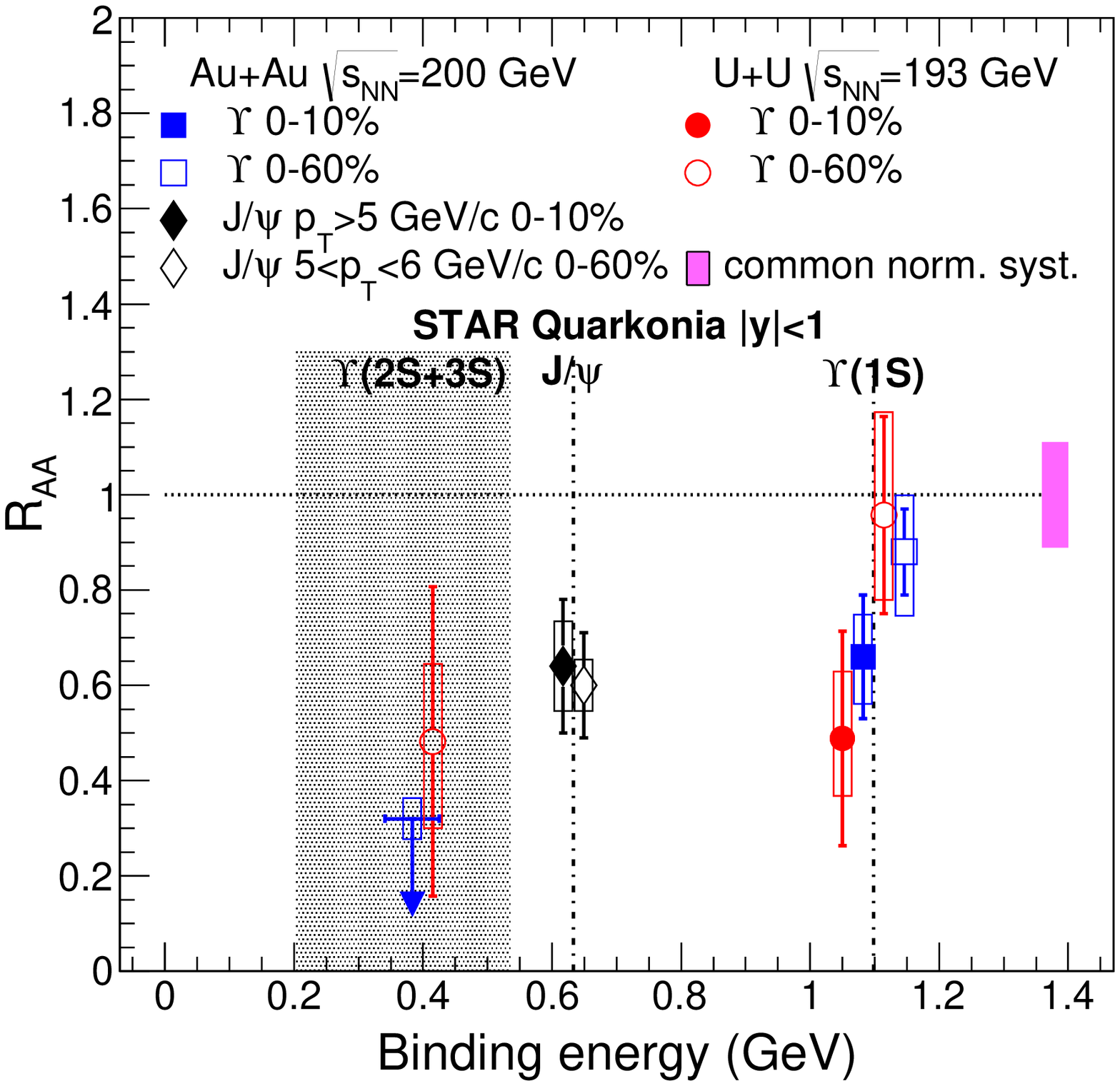}
\caption{\label{fig:raa_binding}\Raa versus binding energy for \Ups(1S) and \Ups{}(2S+3S) in U+U collisions at $\sqsn=193$ GeV~\cite{Adamczyk:2016dzv}, compared to the \Ups and high-\pT \Jpsi results in Au+Au collisions at $\sqsn=200$ GeV~\cite{Adamczyk:2013poh,Adamczyk:2012ey}. The 95\% upper confidence bound is shown for \Ups{}(2S+3S) measured in Au+Au collisions.}
\end{figure}

Figure~\ref{fig:raa_npart} shows the nuclear modification factor of \Ups{}(1S+2S+3S) and \Ups{}(1S), respectively, for U+U collisions at $\sqsn=193$ GeV and Au+Au collisions at $\sqsn=200$ GeV.
The new U+U data consolidate observations made previously in Au+Au data: considering 0-10\% centrality Au+Au and U+U points together, both the \Ups{}(1S+2S+3S) and \Ups{}(1S) points are significantly suppressed, similar in extent to data from the LHC~\cite{Khachatryan:2010zg} but this suppression is not complete~\cite{Adamczyk:2016dzv}. Theory comparisons~\cite{Emerick:2011xu,Strickland:2011aa,Liu:2010ej} show that data favor scenarios where the $q\bar{q}$ pairs are strongly bound, over the weakly bound ones.
Figure~\ref{fig:raa_binding} shows the nuclear modification factor of \Ups states measured in U+U collisions at $\sqsn=193$ GeV and Au+Au collisions at $\sqsn=200$ GeV, compared to high-\pT{} \Jpsi in Au+Au collisions at $\sqsn=200$ GeV~\cite{Adamczyk:2012ey}. The emerging picture supports sequential melting of the quarkonium states with different binding energies. While the U+U data show a non-significant presence of the excited states, it is still consistent with the upper limit established previously in the Au+Au measurement.

\section{Production of \Ups{} in Au+Au collisions with the MTD}

\begin{figure}[h!]
\includegraphics[width=\columnwidth]{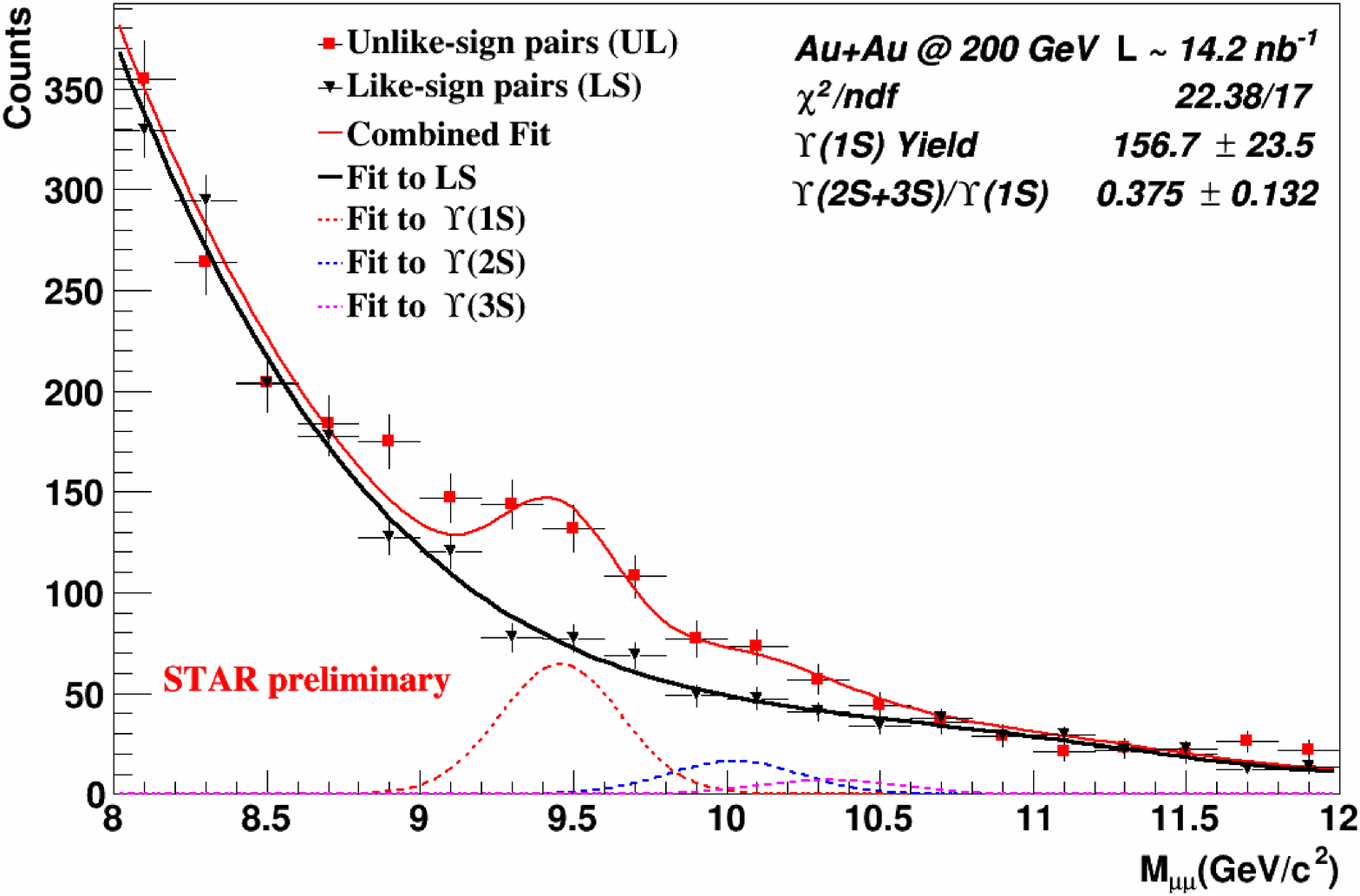}
\caption{\label{fig:mtdmass}
Invariant mass distribution of $\Upsilon\rightarrow\mu^+\mu^-$ candidates and combinatorial background in the 0-80\% centrality class of Au+Au collisions at $\sqsn=200$ GeV.}
\end{figure}

\begin{figure}[h!]
\includegraphics[width=\columnwidth]{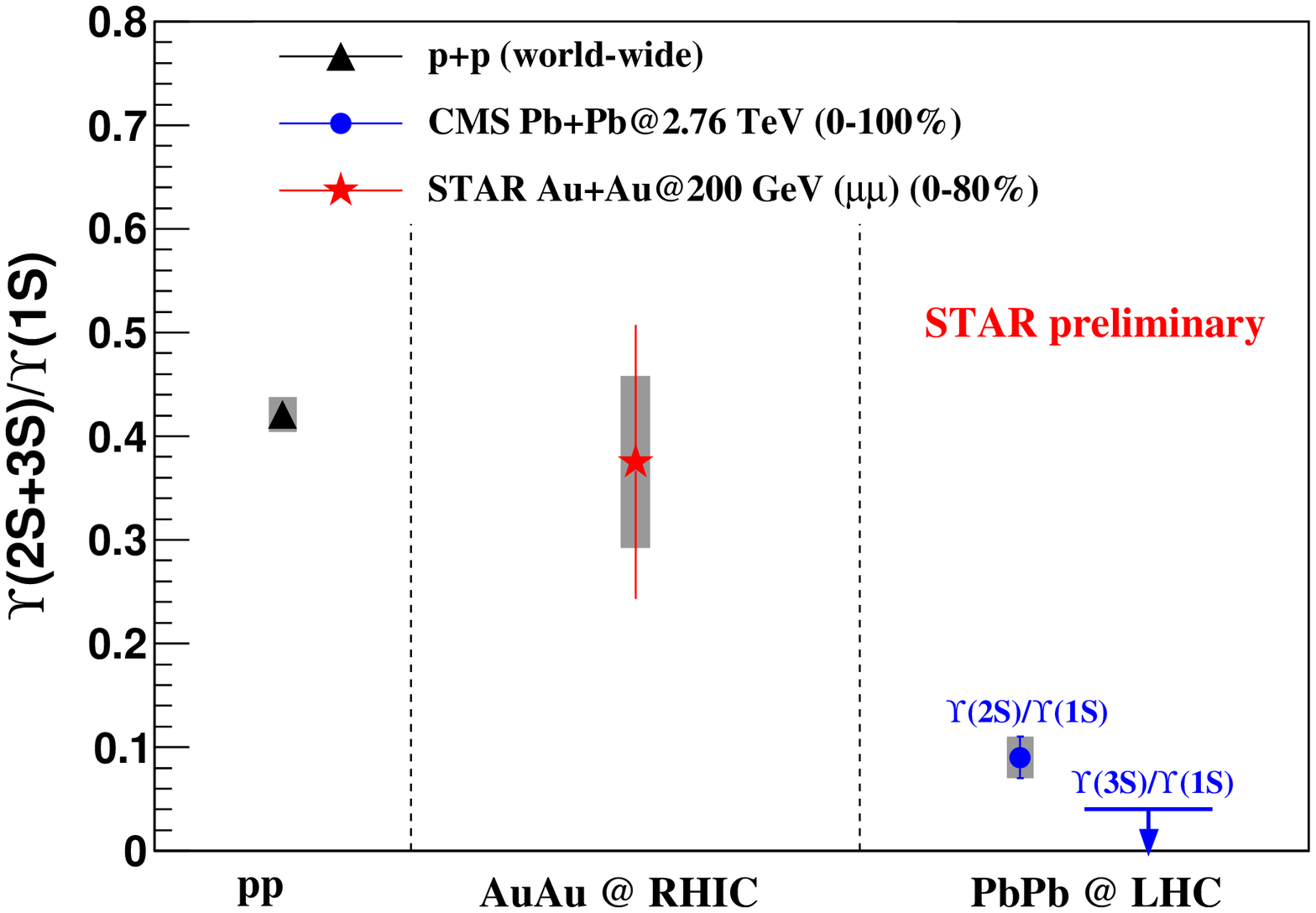}
\caption{\label{fig:mtdratio}Excited-to-ground state ratio of the \Ups mesons via the dimuon channel, compared to CMS measurements~\cite{Chatrchyan:2012lxa}.}
\end{figure}

Measurements in the dielectron channel suffer from the electron bremsstrahlung tail that causes the higher mass states to give contribution to the peak region of lower mass states. The advantage of $\Ups\rightarrow\mu^+\mu^-$ measurements is that muons are less affected by bremsstrahlung, thus it is less of a challenge to separate the different \Ups{} states with comparable statistics.
The Muon Telescope Detector (MTD), completed in 2014, is designed for precision measurements of quarkonia production via the dimuon channel. It is installed outside the solenoidal magnet of STAR, covering 45\% of the azimuth angle within the pseudorapidity range $|\eta|<0.5$. Its Multi-gap Resistive Plate Chamber (MRPC) technology provides means to trigger on muons and identify them.
Figure~\ref{fig:mtdmass} shows the invariant mass spectrum of \Ups candidates reconstructed via the dimuon channel.
Here, as in the dielectron channel, the raw yields of the \Ups states are obtained from a simultaneous fit on the like-sign and unlike-sign muon pairs. The individual \Ups{}($n$S) states are modelled with Gaussian distributions, each centered at the PDG value of the given state. Figure~\ref{fig:mtdratio} shows the excited-to-ground state ratio \Ups{}(2S+3S)/\Ups{}(1S) via the dimuon channel in Au+Au collisions at $\sqsn=200$ GeV.
The result is compared to worldwide data in p+p collisions, as well as measuremens from CMS in Pb+Pb collisions at $\sqsn=2.76$ TeV~\cite{Chatrchyan:2012lxa}. The result suggests that the excited states may not be as strongly suppressed at RHIC as at the LHC, although the errors are large.

\section{Summary and outlook}
Measurements of \Ups{}(1S+2S+3S) and \Ups{}(1S) production in U+U collisions at $\sqsn=193$ GeV consolidate the trends observed in Au+Au collisions and extend them toward higher \Npart values. 
We observe a significant suppression of the \Ups(1S+2S+3S){} signal in heavy-ion collisions at RHIC top energies. The suppression of the \Ups{}(1S) state in central heavy-ion collisions is also confirmed by the new U+U data. New preliminary measurements of \Ups production via the dimuon decay channel show an indication that the excited  \Ups{}(2S+3S) states are not completely suppressed in Au+Au collisions at $\sqsn=200$ GeV, and hint that the dissociation of the excited \Ups{}(2S+3S) states is less prominent at RHIC than at LHC energies. 
There are ongoing analyses of the full 2014--2016 dataset of Au+Au collisions in both the dimuon and dielectron channels. Once completed, these will provide more precise information about the extent of excited state suppression. In parallel to that, analysis of p+A data recorded in 2015 will confirm or exclude subtantial CNM effects at mid-rapidity.


This work has been supported by the Hungarian NKFIH/OTKA NK 106119 and K 120660 grants, the J\'anos Bolyai scholarship of the Hungarian Academy of Sciences, as well as the grant 13-20841S of the Czech Science Foundation (GA\v{C}R).

\end{document}